\documentclass[a4paper,12pt, epsfig]{article}  \pdfoutput=1
\usepackage{epsfig}
\usepackage{amssymb,latexsym}
\usepackage{amsfonts}
\usepackage{amsmath}

\newskip\humongous \humongous=0pt plus 1000pt minus 1000pt

\newif\ifdtup

\catcode`\@=11

\@addtoreset{equation}{section}
\def\theequation{\thesection.\arabic{equation}}

\def\@normalsize{\@setsize\normalsize{15pt}\xiipt\@xiipt
\abovedisplayskip 14pt plus3pt minus3pt%
\belowdisplayskip \abovedisplayskip
\abovedisplayshortskip \z@ plus3pt%
\belowdisplayshortskip 7pt plus3.5pt minus0pt}

\def\small{\@setsize\small{13.6pt}\xipt\@xipt
\abovedisplayskip 13pt plus3pt minus3pt%
\belowdisplayskip \abovedisplayskip
\abovedisplayshortskip \z@ plus3pt%
\belowdisplayshortskip 7pt plus3.5pt minus0pt
\def\@listi{\parsep 4.5pt plus 2pt minus 1pt
      \itemsep \parsep
      \topsep 9pt plus 3pt minus 3pt}}

\relax

\catcode`@=12

\catcode`\@=11

\def\section{\@startsection{section}{1}{\z@}{3.5ex plus 1ex minus
    .2ex}{2.3ex plus .2ex}{\large\bf}}

\def\thesection{\arabic{section}}
\def\thesubsection{\arabic{section}.\arabic{subsection}}

\def\appendix{\setcounter{section}{0}
  \def\thesection{Appendix \Alph{section}}
  \def\thesubsection{\Alph{section}.\arabic{subsection}}
  \def\theequation{\Alph{section}.\arabic{equation}}}

\def\SymBoxes#1#2#3#4{\newdimen\un@t \un@t#3%
\raisebox{#1}{\rule{#2\un@t}{#4}\hskip-#2\un@t
\@tempdimb\un@t \advance\@tempdimb by-#4\@tempcntb#2\relax%
\@whilenum{\@tempcntb>0}\do{
\rule{#4}{\un@t}\hskip\@tempdimb \advance\@tempcntb by\m@ne}%
\hskip-#2\un@t \rule[\un@t]{#2\un@t}{#4}%
\rule[\un@t]{#4}{#4}\hskip-#4
\rule{#4}{\un@t}}\hskip-#4}                

\begin{document}


\newcommand{\dd}{\textrm{d}}

\newcommand{\beq}{\begin{equation}}
\newcommand{\eeq}{\end{equation}}
\newcommand{\bea}{\begin{eqnarray}}
\newcommand{\eea}{\end{eqnarray}}
\newcommand{\beas}{\begin{eqnarray*}}
\newcommand{\eeas}{\end{eqnarray*}}
\newcommand{\defi}{\stackrel{\rm def}{=}}
\newcommand{\non}{\nonumber}
\newcommand{\bquo}{\begin{quote}}
\newcommand{\enqu}{\end{quote}}
\renewcommand{\(}{\begin{equation}}
\renewcommand{\)}{\end{equation}}
\def\de{\partial}
\def\Om{\ensuremath{\Omega}}
\def\Tr{ \hbox{\rm Tr}}
\def\rc{ \hbox{$r_{\rm c}$}}
\def\H{ \hbox{\rm H}}
\def\HE{ \hbox{$\rm H^{even}$}}
\def\HO{ \hbox{$\rm H^{odd}$}}
\def\HEO{ \hbox{$\rm H^{even/odd}$}}
\def\HOE{ \hbox{$\rm H^{odd/even}$}}
\def\HHO{ \hbox{$\rm H_H^{odd}$}}
\def\HHEO{ \hbox{$\rm H_H^{even/odd}$}}
\def\HHOE{ \hbox{$\rm H_H^{odd/even}$}}
\def\K{ \hbox{\rm K}}
\def\Im{ \hbox{\rm Im}}
\def\Ker{ \hbox{\rm Ker}}
\def\const{\hbox {\rm const.}}
\def\o{\over}
\def\im{\hbox{\rm Im}}
\def\re{\hbox{\rm Re}}
\def\bra{\langle}\def\ket{\rangle}
\def\Arg{\hbox {\rm Arg}}
\def\exo{\hbox {\rm exp}}
\def\diag{\hbox{\rm diag}}
\def\longvert{{\rule[-2mm]{0.1mm}{7mm}}\,}
\def\a{\alpha}
\def\b{\beta}
\def\e{\epsilon}
\def\l{\lambda}
\def\ol{{\overline{\lambda}}}
\def\ochi{{\overline{\chi}}}
\def\th{\theta}
\def\s{\sigma}
\def\oth{\overline{\theta}}
\def\ad{{\dot{\alpha}}}
\def\bd{{\dot{\beta}}}
\def\oD{\overline{D}}
\def\opsi{\overline{\psi}}
\def\dag{{}^{\dagger}}
\def\tq{{\widetilde q}}
\def\L{{\mathcal{L}}}
\def\p{{}^{\prime}}
\def\W{W}
\def\N{{\cal N}}
\def\hsp{,\hspace{.7cm}}
\def\bo{\ensuremath{\hat{b}_1}}
\def\bfo{\ensuremath{\hat{b}_4}}
\def\co{\ensuremath{\hat{c}_1}}
\def\cfo{\ensuremath{\hat{c}_4}}
\newcommand{\C}{\ensuremath{\mathbb C}}
\newcommand{\Z}{\ensuremath{\mathbb Z}}
\newcommand{\R}{\ensuremath{\mathbb R}}
\newcommand{\rp}{\ensuremath{\mathbb {RP}}}
\newcommand{\cp}{\ensuremath{\mathbb {CP}}}
\newcommand{\vac}{\ensuremath{|0\rangle}}
\newcommand{\vact}{\ensuremath{|00\rangle}                    }
\newcommand{\oc}{\ensuremath{\overline{c}}}

\newcommand{\Vol}{\textrm{Vol}}

\newcommand{\half}{\frac{1}{2}}

\def\changed#1{{\bf #1}}

\begin{titlepage}

\def\thefootnote{\fnsymbol{footnote}}

\begin{center}
{\large {\bf
Stability of Closed Timelike Curves in a Galileon Model
  } }

\bigskip

{\large
\noindent
Jarah Evslin\footnote{\texttt{jarah@ihep.ac.cn}}}
\end{center}

\renewcommand{\thefootnote}{\arabic{footnote}}

\begin{center}
\vspace{0em}
{\em  { 
Theoretical Physics Center for Science Facilities\\
Institute of High Energy Physics\\
Chinese Academy of Sciences\\
YuQuan Lu 19(B)\\Beijing 100049\\ China\\

\vspace{.2cm}


\vskip .4cm}}

\end{center}

\vspace{3.1cm}

\noindent
\begin{center} {\bf Abstract} \end{center}

\noindent
Recently Burrage, de Rham, Heisenberg and Tolley have constructed eternal, classical solutions with closed timelike curves (CTCs) in a Galileon model coupled to an auxiliary scalar field.  These theories contain at least two distinct metrics and, in configurations with CTCs, two distinct notions of locality.  As usual, globally CTCs lead to pathologies including nonlocal constraints on the initial Cauchy data.  Locally, with respect to the gravitational metric, we use a WKB approximation to explicitly construct small, short-wavelength perturbations without imposing the nonlocal constraints and observe that these perturbations do not grow and so do not lead to an instability.

\vfill

\begin{flushleft}
{\today}
\end{flushleft}
\end{titlepage}

\hfill{}


\setcounter{footnote}{0}

\section{The goal}
The central question in physics is:  Given a configuration on an initial surface, local rules for evolving it to later times and boundary conditions at large spatial distances, what is the configuration at a future time?  Of course in practice one can only know the initial conditions up to some bounded uncertainty.  A theory is clearly only useful if this finite uncertainty in the initial conditions leads to a finite uncertainty in the results.  In the case of most classical field theories, the rules for evolution are hyperbolic partial differential equations, and so it suffices that the initial surface be a spatial Cauchy surface.  Generically small perturbations are described by the wave equation, and so small high wavenumber perturbations of the initial conditions lead to high frequency perturbations of the time evolution which, critically, have the same amplitude as the original perturbations and so remain finite.  If on the other hand one tries to place initial data on a timelike surface and then evolve, small but arbitrarily high wavenumber perturbations in the initial conditions lead to arbitrarily large perturbations after a fixed amount of evolution, therefore initial conditions placed on such surfaces cannot be used to determine the configuration elsewhere.

In a universe with closed timelike curves (CTCs) there is an additional constraint.  Evolving the surface into the future using local equations of motion, one can find the configurations on future surfaces.  But iterating this process, one eventually returns to the initial surface itself.  The physical observables must be single-valued, which means that the new values on the surface must be equal to the initial conditions.  This constraint is in generally nonlocally encoded in the initial data, and so it presents a serious if not insurmountable complication to the theory.  One may be tempted to simply define an initial surface to which the CTCs are always parallel.  Then these constraints can be satisfied by simply imposing a periodicity condition on the surface, such that circumnavigating the curve one arrives at the same values.  However, as the curve is timelike, the surface is not spacelike and so again cannot be integrated to give the configuration elsewhere up to any finite error.  If one tries to go ahead and try to integrate the equations of motion starting from initial data on such a surface, one will find a wrong sign dispersion relation leading to an apparently complex energy for the plane wave at high momenta, which indicates an instantly fatal instability.

Does this apparent instability indicate that the configuration is really unstable, and so perhaps the CTC does not develop?  Any configuration, even a free theory in Minkowski space, exhibits such an instability if one attempts to integrate initial conditions defined on a timelike slice.  Therefore, on its own, this feature does not imply an instability.  On the contrary, if the equations are hyperbolic then a spacelike Cauchy surface can be constructed and its data integrated, leading to no obvious pathologies in the classical theory apart from the aforementioned nonlocal constraint.

In this short note we would like to argue that the classical instability discovered in Ref.~\cite{clare}, in a variant of the Galileon model\label{Galileon}, is of this kind and so does not necessarily indicate that the classical theory cannot develop CTCs.  While the authors did not directly impose initial conditions on a timelike slice, the periodicity condition which they use to solve the nonlocal constraints effectively fixes initial conditions for each mode.   Physical high frequency instabilities occur when the timelike eigenvalue in the effective metric crosses zero and ghostlike instabilities when all eigenvalues change signs.  On the other hand, in this case we will show that no eigenvalue of either relevant metric changes sign. 

The difference in points of view may appear to be only be semantic:  On the one hand in Ref.~\cite{clare}, the authors observed that a small perturbation which obeys the nonlocal constraints leads to an instantaneous instability and so invalidates the solution.  On the other hand in this note, using only spatial Cauchy surfaces, we will see in Sec.~\ref{ultsez} that an infinitesimal deformation propagates with a constant amplitude, but most likely is inconsistent with the nonlocal constraints and so again invalidates the solution.  However, there is a distinction between these two pathologies.  In Ref.~\cite{clare} it was claimed that the instability means that the theory avoids the CTC and so is consistent.  The nonlocal constraint, on the other hand, yields a Galileon field which is generically is multiply defined and so the classical theory is inconsistent.  Of course, if the Galileon is embedded in a healthy UV completion, this second problem is simply a sign that the low energy effective theory breaks down, but we will claim that this breakdown is not evident in any of the data within a local neighborhood where distances are determined using the nondynamical gravitational metric.  Clearly an effective theory is more useful if there is a local criterion which determines when it can and cannot be trusted.

\section{History of CTCs in Galileon models}

The Galileon model is the most general theory of an interacting scalar field $\phi$ with a particular global symmetry under shifts and also linear transformations and with equations of motion that only depend upon the first two derivatives of $\phi$, guaranteeing the absence of Ostrogradski ghosts.  The Lagrangian density is
\beq
\L=c_1\L_1-\frac{1}{2}\partial_\mu\phi\partial^\mu\phi+c_3\L_3+c_4\L_4+c_5\L_5 \label{Lag}
\eeq
where $c_k$ are real parameters and the $\L_k$ are given in Ref.~\cite{Galileon}.   For simplicity we will consider the special case $c_1=c_4=c_5=0$ which describes a limit of the 5-dimensional DGP model \cite{DGP}.   The only term $\L_k$ that we will then need is
\beq
\L_3=-\frac{1}{2}(\Box\phi)(\partial_\nu\phi)^2. \label{nonlin}
\eeq
The Galileon model is interesting in part because it contains ghost-free and classically stable solutions which violate the null energy condition \cite{Galileon} and even healthy solutions which cross from a domain in which they do not violate this condition into a domain in which they do  \cite{balza} realizing the quintom scenario proposed for example in Ref.~\cite{quintom}.  Recently it has been shown that if the Galileon symmetry is gauged, then the Galileon is the helicity zero component of a massive graviton \cite{massive}, expanding upon the identification of the corresponding kinetic term in \cite{nimagrav}.

Both the DGP \cite{consistency} and Galileon \cite{positivity} models exhibit superluminal propagation.  More precisely, small perturbations
\beq
\phi=\phi_0+\delta\phi. \label{pdicomp}
\eeq
about a nontrivial background field configuration $\phi_0$ are described by the wave equation in a background with an effective metric which differs from the metric of spacetime
\beq
\delta\L=-\frac{1}{2}(\partial_\mu\phi)^2(1+2c_3\Box\phi_0)+c_3(\partial^\mu\partial^\nu\phi_0)\partial_\mu\delta\phi\partial_\nu\delta\phi
=-\frac{1}{2}G^{\mu\nu}\partial_\mu\delta\phi\partial_\nu\delta\phi.
\eeq
If $\phi_0$ is harmonic with respect to the Minkowski gravitational metric then, normalizing $\phi$ so that $c_3=1$,  the inverse effective metric is just
\beq
G^{\mu\nu}=\eta^{\mu\nu}-\partial^\mu\partial^\nu\phi_0 .
\eeq
The scalar field propagates along null directions with respect to the effective metric, which in generic configurations can be timelike or spacelike with respect to the gravitational metric.  Such superluminal propagation is not necessarily inconsistent so long as it does not lead to closed timelike curves \cite{Alexribut}, and in fact has recently become quite popular in phenomenology \cite{posdep}.  Nonetheless, it has been proposed in Ref.~\cite{Nima} that such superluminal propagation may be used to create configurations with CTCs, along the lines suggested in Ref.~\cite{prenima}.   

This proposal was followed in Ref.~\cite{taotao} where the authors constructed a classical Galileon solution, in the DGP case, which they claim is stable and develops CTCs.  The construction began from the observation that left-moving plane wave solutions 
\beq
\phi_0=f(x+t)
\eeq
allow luminal propagation for right-moving Galileons and superluminal propagation for left-movers.  More precisely, Galileons travelling in the same spatial direction as the plane wave travel along the direction proportional to $(f^{\prime\prime}+1)\hat{t}+(f^{\prime\prime}-1)\hat{x}$ which is null with respect to the effective metric
\beq
G_{\mu\nu}= \left(
\begin{array}{cc}
  1-f^{\prime\prime} & f^{\prime\prime}  \\
 f^{\prime\prime}  & -1-f^{\prime\prime} 
\end{array}
\right) 
\eeq
but superluminal with respect to the gravitational metric if $f^{\prime\prime}$ is negative.  In fact, if $f^{\prime\prime}<-1$ then the Galileon travels backwards in time, which is a necessary condition at least somewhere in order to create a CTC.  For simplicity the authors considered boundary conditions in which the Galileon and its first derivative vanish at both ends of straight rods, which means that $f^{\prime\prime}$ cannot always be negative, as it must integrate to zero.  The positivity of $f^{\prime\prime}$ would lead to a massive subluminality which would eliminate the CTC.  To avoid this problem, the authors included another plane wave moving to the right, for which the positive $f^{\prime\prime}$ could be added without imposing subluminality on Galileons traveling to the left, since Galileons traveling in the opposite direction to the plane wave are always exactly luminal.  The solution was thus complicated, but appeared consistent and stable.  In fact, locally the effective metric was that of flat space \cite{clare}.

In Ref.~\cite{clare} the previous construction was simplified by considering a single plane wave traveling around a compact circle.  In this case, the well-definedness of the Galileon field clearly requires its second derivative to integrate to zero, as in the case of the previous paper.  With a single plane wave, this leads to a very subluminal Galileon inside of the region in which the second derivative is positive.  The authors avoided this problem by coupling the Galileon $\phi$ to an additional field $\chi$ which is sensitive to the gravitational metric $\eta$
\beq
\L
\supset -\frac{1}{2}G^{\mu\nu}\partial_\mu\delta\phi\partial_\nu\delta\phi-\frac{1}{2}\eta^{\mu\nu}\partial_\mu\chi\partial_\nu\chi-\delta\phi\hat{\mu}\chi
\eeq
where $\hat{\mu}$ is a kinetic operator consisting of various contractions of derivatives, they argued that such couplings are generated in a variety of settings.  The field $\chi$ always travels luminally and so can bypass the region in which the Galileon field would be constrained to be subluminal without losing too much time to close the CTC.  This CTC therefore can be circumnavigated only by two fields which each feel different metrics, motivating their name ``bimetric CTCs.'' 

Generalizing the usual arguments of Hawking's chronology protection conjecture \cite{hawking} the authors claimed that in the quantum theory, in a free field truncation, one expects a divergent backreaction which, depending on its sign, may inhibit the formation of CTCs.  However they also claimed that the system would not arrive at this point, because already classically under arbitrarily small perturbations with wavenumber $k$ it decays in a characteristic time $1/k$, which for sufficiently high $k$ is instantaneous.  Of course if the CTCs do form, then this $k$ is an initial condition on a temporal surface, and so the instability is just a result of the fact that the initial conditions were not defined on a spacelike surface.  Before the formation of the curve there is no such instability but the energy measured with respect to the timelike isometry diverges.  This should come as no surprise, as the corresponding Killing vector becomes lightlike as the CTCs (null Cauchy surface) form and so the corresponding frame is effectively infinitely boosted.  In the next section we will review these solutions and determine their responses to such perturbations.  We will see that there is always a choice of time direction such that these apparent local instabilities and divergences are absent from the classical theory.

\section{Bimetric CTCs} \label{ultsez}

\subsection{Perturbations}

We will now review the bimetric CTCs of Ref.~\cite{clare} and study their classical stability with respect to small, local, high wavenumber perturbations.  As we are interested in high wavenumber perturbations, a WKB approximation will be considered in which the magnitudes of the fields $\delta\phi$ and $\chi$ change slowly, and so the differential operators act to leading order on their phases.  More precisely, we will write these fields locally in the plane wave form
\beq
\delta\phi=\Phi e^{i\vec{k}\cdot \vec{x}-ik_t t}\hsp
\chi=X e^{i\vec{k}\cdot \vec{x}-ik_t t} \label{wkb}
\eeq
with the approximation that the derivatives of the wavenumbers $k$ are smaller than the various products $k^2$.  

Now the equations of motion are simply the algebraic condition that the vector $(\Phi,X)$ be annihilated by a matrix determined in terms of the functions $\vec{k}$ and $k_t$.  Following \cite{clare} it is convenient to introduce the light cone momenta
\beq
k_\pm=k_x\pm k_t\hsp p^2=k_y^2+k_z^2.
\eeq
Then the algebraic equations of motion, to leading order in the WKB approximation, are
\beq
\left(
\begin{array}{cc}
  -k_+^2f^{\prime\prime}+p^2+k_+k_-&\mu\dag  \\
 \mu  & p^2+k_+k_-
\end{array}
\right) 
\left(
\begin{array}{c}
 \Phi \\
 X
\end{array}
\right)=0 . \label{matrice}
\eeq
The solutions to this equation correspond to values of the momenta for which the determinant of the matrix vanishes.  This degeneracy condition can be solved for $k_-$ 
\beq
k_-^{(\pm)}=\frac{1}{2k_+}\left(-2p^2+k_+^2f^{\prime\prime}\pm\sqrt{k_+^4f^{\prime\prime 2}+4\mu\mu\dag}\right).\label{radici}
\eeq

Recall that while $f^{\prime\prime}$ on average is zero, it is less than $-1$ while the Galileon travels backwards in time, which is a necessary condition for the formation of a CTC.  Therefore there will be regions in which $|f^{\prime\prime}|\geq 1$.  The construction of Ref.~\cite{clare} assumes that, at least inside of these regions
\beq
4\mu\mu\dag<< k_+^4
\eeq
and so in these regions one may approximate
\beq
k_-^{(+)}=\left\{
\begin{array}{cl}
\frac{p^2}{k_+}+k_+f^{\prime\prime}&{\rm{if}}\ f^{\prime\prime}\geq 0\\
\frac{p^2}{k_+}&{\rm{if}}\ f^{\prime\prime}\leq 0
\end{array}
\right. \hsp
k_-^{(-)}=\left\{
\begin{array}{cl}
\frac{p^2}{k_+}&{\rm{if}}\ f^{\prime\prime}\geq 0\\
\frac{p^2}{k_+}+k_+f^{\prime\prime}&{\rm{if}}\ f^{\prime\prime}\leq 0
\end{array}
\right. . \label{ksol}
\eeq
Notice that, if $f^{\prime\prime}\geq 0$, then $k_-^{(+)}$ solves (\ref{matrice}) in the case $X=\mu=0$ and $k_-^{(-)}$ in the case  $\Phi=\mu=0$.  In other words, to leading order in $\mu^2/k_+^4$, the solution $k_-^{(+)}$ describes a purely Galileon perturbation in the region in which $f^{\prime\prime}\geq 0$ and a purely $\chi$ perturbation in the region in which $f^{\prime\prime}\leq 0$.  Recall that the Galileon is subluminal in the regime in which $f^{\prime\prime}\geq 0$, while $\chi$ is always exactly luminal.  Therefore the $k_-^{(+)}$ fluctuations are never superluminal.  This is the reason that no instability was found in Ref.~\cite{clare} when initial conditions were chosen for $k_-^{(+)}$ on the CTC, because this curve is in fact spacelike with respect to the metric felt by $k_-^{(+)}$ and so the Cauchy problem is well defined.  On the other hand, $k_-^{(-)}$ is nearly all Galileon in the regime $f^{\prime\prime}\leq 0$, where the Galileon is superluminal, and is nearly all $\chi$ where the Galileon is subluminal.  Therefore these are the modes which can circumnavigate the CTCs, and so those that were problematic in Ref.~\cite{clare}.

Between the regions in which $f^{\prime\prime}$ is positive and negative, it of course must have a zero.  In these regions the $\mu^2$ term dominates the square root in (\ref{radici}) and so Eq.~(\ref{ksol}) does not apply.  In these regions, each solution interpolates between the Galileon dominated and $\chi$ dominated regimes.  This explains the observation in Ref.~\cite{clare} that even for arbitrarily small $\mu$, the behavior of these solutions does not reproduce that of the uncoupled case, in which no interpolation occurs.  Any small, but nondegenerate, kinetic operator will necessarily dominate in the square root near the zeros of $f^{\prime\prime}$, leading to an interpolation which becomes significant as $f^{\prime\prime}$ again becomes large.

\subsection{Stability}

As $k_-^{(+)}$ perturbations are not superluminal, and were shown to be healthy in Ref.~\cite{clare}, we will not consider them further.  The authors found that, when fixing a wavenumber for $k_-^{(-)}$ on the closed timelike curve, the evolution in another direction is unstable with respect to $k_y$ fluctuations.  However, once the CTC has formed this curve is timelike for the $k_-^{(-)}$ perturbations, and so fixing the wavenumber along the curve leaves an elliptic differential equation for the $k_y$ and perpendicular spatial directions.  Elliptic differential equations cannot be integrated from a single initial slice due to just this instability.  Such a pathology would be found if one imposed initial conditions on a timelike surface in any spacetime, and so on its own does not indicate a physical instability.  In addition the divergent energy that they found as the closed timelike curves adiabatically form are seen in any background if one defines the energy with respect to a translation which is rotated from timelike to null.

While a general search for instabilities is beyond the scope of this work, we will search for high frequency, classical, local instabilities.  These depend only upon the local information available, in particular on the local value of the effective metric.  High frequency instabilities occur when the effective metric becomes Euclidean, whereas ghost instabilities occur when it changes sign.  We will show that no eigenvalues change sign.

The two different mass-shell conditions in Eq.~(\ref{ksol}) indicate two different effective metrics, one in the Galileon dominated regime $f^{\prime\prime}\leq 0$ and one in the $\chi$ dominated regime $f^{\prime\prime}\geq 0$.  By construction the $\chi$ field is sensitive to the Minkowski metric, whose eigenvalues are constant and so no local instability is possible.  Therefore we will focus on perturbations in the Galileon regime.  

The mass shell condition (\ref{ksol}) indicates that the Galileon field is sensitive to the same inverse metric as in Ref.~\cite{taotao}
\beq
G^{\mu\nu}= \left(
\begin{array}{cccc}
  1+f^{\prime\prime} & f^{\prime\prime} &0&0\\
 f^{\prime\prime} & -1+f^{\prime\prime}&0&0\\
0&0&-1&0\\
0&0&0&-1
\end{array}
\right) .
\eeq
The determinant of this metric is always equal to $-1$, for any value of $f^{\prime\prime}$, therefore no eigenvalues can change sign and no instability occurs.  

The unstable modes discovered in Ref.~\cite{clare} are waves traveling in the $y$ direction.  The authors imposed boundary conditions on the closed timelike curve, which quantized the $k_x$ component.  They found a high frequency instability for every value of $k_x$, so for simplicity we will set $k_x=0$ for the moment.  In this case, a wave in the $t-y$ plane feels the diagonal effective metric with values $1+f^{\prime\prime}$ and $-1$.  Clearly, when $f^{\prime\prime}<-1$, as it must be for the Galileon to travel backwards in time, this metric becomes Euclidean and a high frequency instability occurs.  Equivalently, the energy computed using the derivative with respect to the $t$ direction is complex when $+f^{\prime\prime}<-1$ and diverges as $+f^{\prime\prime}$ approaches $-1$.

However, this instability relies critically on fixing $k_x$, which corresponds to fixing an initial condition on the timelike surface.  In fact, the $x$ direction is timelike precisely when  $f^{\prime\prime}<-1$.   What happens if one does not fix $k_x$?  Is there a stable oscillation mode for any fixed value of $k_y$?  In other words, if one perturbs the Galileon field locally with a wave which rapidly oscillates in the $y$ direction and lets it evolve, what happens to the amplitude of the oscillation?  Of course, to define this question, one must define evolve.  Locally, evolve means to integrate the equations of motion in a temporal direction, that is a direction whose norm with respect to the effective metric is positive.  In our WKB approximation this is equivalent to solving the mass shell conditions for $\omega$, where $\omega$ is the frequency with respect to a positive-normed direction.

One direction whose norm is certainly positive is the timelike eigenvector of the inverse effective metric
\beq
(k_t,k_x)=(f^{\prime\prime},-1+\sqrt{f^{\prime\prime^2}+1}).
\eeq
Therefore the mass-shell condition
\beq
0=k_\mu G^{\mu\nu}k_\nu=(1+f^{\prime\prime})k_t^2+2f^{\prime\prime}k_tk_x+(-1+f^{\prime\prime})k_x^2+k_y^2+k_z^2
\eeq
 may be satisfied for a high frequency perturbation which oscillates in the $y$ direction with an arbitrary momentum $k_y$ by simply adding to $k$ a multiple of this timelike eigenvector
\beq
k\hspace{-.1cm} =\hspace{-.1cm} \left(\frac{-k_yf^{\prime\prime}}{\sqrt{2(f^{\prime\prime 2}-f^{\prime\prime}+1)(f^{\prime\prime}+\sqrt{f^{\prime\prime 2}+1})}},\frac{k_y(1-\sqrt{f^{\prime\prime2}+1})}{\sqrt{2(f^{\prime\prime 2}-f^{\prime\prime}+1)(f^{\prime\prime}+\sqrt{f^{\prime\prime 2}+1})}},k_y,0\right). \label{finale}
\eeq
This wavevector, in the WKB approximation (\ref{wkb}), describes the classical evolution of a local, high-frequency Galileon perturbation which oscillates in the $y$ direction with arbitrary wavenumber $k_y$.  The positivity of the expressions in the various square roots of (\ref{finale}) imply that all components of $k$ are real, and so no local instability is present for any value of $f^{\prime\prime}$, at least to leading order in the WKB approximation.  

Of course the vanishing proper length of this curve at the moment at which it becomes null means that another notion of locality exists for the Galileon field, using the effective metric, in which the entire curve is local\footnote{More precisely, for any $\epsilon>0$ the entire curve is within an open ball of radius $\epsilon$ centered on any point on the curve.}.  Instabilities which are local in that sense are considerably more difficult to analyze, resembling the situation with CTCs in gravity.  As described in Refs. \cite{hawking,wald,clare} the vanishing proper distance may indicate a divergent stress tensor in the quantum theory, which certainly would lead to a large backreaction in the gravitational theory and perhaps also in this nongravitational theory.  However the sign of this backreaction is important.  It may be that it prohibits the formation of the CTC as was speculated in the case of free fields in Refs \cite{hawking,Horava,clare}, or it may be that the UV theory intervenes only to cut off the divergence without inhibiting the formation of CTCs~\cite{kip}.

\section* {Acknowledgement}

\noindent
JE is supported by the Chinese Academy of Sciences Fellowship for Young International Scientists grant number 2010Y2JA01.   He is honored to express his profound gratitude to Andrew Tolley for illuminating correspondence.


\end{document}